
\magnification=1200
\hsize=13cm


\def\ep{e^{2i\pi}}

\let \part\partial

\def\o#1#2{{#1\over#2}}

\def\un{{\bf 1}}
\def\sqd{\sqrt{2}}

\def\twomat#1#2#3#4{\left(\matrix{#1&#2\cr #3&#4\cr}\right)}
\def\twovec#1#2{\left(\matrix{#1\cr #2\cr}\right)}


\def\ep4{e^{i \pi /4}}

\def\emp4{e^{-i \o{\pi}{4}}}

\def\twomat#1#2#3#4{\left(\matrix{#1&#2\cr #3&#4\cr}\right)}
\def\twovec#1#2{\left(\matrix{#1\cr #2\cr}\right)}

\def\e2pi{e^{2\pi i}}
\def\em2pi{e^{- 2\pi i}}

\def\eip2n{e^{i\pi n/2}}




\def\ep{e^{2i\pi}}

\let \part\partial

\def\e2pi{e^{2\pi i}}
\def\em2pi{e^{- 2\pi i}}


\def\um{\o {1}{2}}

\def\o#1#2{{{#1}\over{#2}}}

\def\un{{\bf 1}}
\def\sqd{\sqrt{2}}

\def\twomat#1#2#3#4{\left(\matrix{{#1}&{#2}\cr {#3}&{#4}\cr}
\right)}
\def\twovec#1#2{\left(\matrix{{#1}\cr {#2}\cr}\right)}

\def\ep4{e^{i \o{\pi}{4}}}
\def\emp4{e^{-i \o{\pi}{4}}}






























\def\eq{\,=\,}

\def\um{{\scriptstyle {\o{1}{2}}}}

\def\ep{e^{+}}
\def\em{e^{-}}

\def\tisei{{\cal M}_{3.3}}
\def\tis{T^6/Z_3}

\centerline{{\bf 1. Introduction}}
\vskip 0.2cm
The list [1] of homogeneous symmetric special K\"ahler manifolds $
\cal S$ [2--6] contains two infinite series $CP_{n-1,1}=SU(1,n)/SU(n)
\otimes U(1)$, $SK(n+1)=SU(1,1)/U(1) \otimes SO(2,n)/SO(2) \otimes SO(n)$
and four exceptional cases including in particular the manifold:
$$
{\cal M}_{3.3}={SU(3,3)/SU(3) \otimes SU(3) \otimes  U(1)}. \eqno(1.1)
$$
In a recent paper [7], the special geometry of the $SK(n+1)$ manifolds
and the associated $SL(2,Z) \times SO(2,n,Z)$ automorphic
superpotentials have been constructed for any $n$. In the present
letter we extend this analysis to the case of the manifold ${\cal M}_{3,
3}$, that is the Teichmuller covering of the moduli space for the
$T^6 /Z_3$ orbifold. In this way we exhaust the analysis of automorphic
superpotentials for the $Z$ orbifold compactification of superstrings.
Indeed, comparing with the list appearing in ref. [8] we see that, with
the exception of $T^6 /Z_3$, all the other (untwisted) orbifold moduli
spaces correspond to special values of $n$ in the $SK(n+1)$ series.

Our goal is to exhibit the special geometry of ${\cal M}_{3.3}$ and to
construct the appropriate infinite sum defining the $SU(3,3,Z)$
automorphic superpotential.

As for any other special manifold $\cal S$, the special geometry of
${\cal M}_{3,3}$ is encoded in a homogeneous of degree two holomorphic
function $F(X)$.
More precisely, for any $\cal S$, we can
costruct a (holomorphic) section $\Omega \equiv (X^\Lambda, i
\partial_\Lambda F(X))$ [4--6] of the symplectic $Sp(2\, dim{\cal
S} +2, R)$ bundle over $\cal S$, from which we extract the K\"ahler
geometry of $\cal S$. In particular the K\"ahler
potential is given by:
$$
G = - \hbox{log} \parallel \Omega \parallel^2 \,\equiv  \,-log \left(
-i\langle ~{\bar \Omega} | \Omega ~ \rangle \right )
 \equiv \, - \log
\left ( {\bar X}^{\Lambda} \partial_{\Lambda} F \, + \,
{\bar \partial}_{\Lambda}{\bar F} X^{\Lambda} \right),
\eqno (1.2)
$$
where
$$
\langle ~\bar \Omega |\Omega ~\rangle=\Omega^\dagger \twomat{0}{\un}
{- \un}{0} \Omega
$$
is the norm of the symplectic section.
The F function proposed in ref. [2],
via supergra\-vity considerations, for the case of ${\cal M}_{3.3}$
is of the
form:
$$
F(X) \sim i \o {\hbox{det} X}{X^0}, \eqno(1.3)
$$
where $X$ is a three by three matrix. Equation (1.3) is just a particular
case of the general formula $F(X)=i d_{\Lambda \Sigma \Delta}
\o{X^{\Lambda} X^\Sigma X^\Delta}{X^0}$, where the $d_{\Lambda\Sigma
\Delta}$ are constant coefficients, valid for any $\cal S$ in the list
of homogeneous symmetric special K\"ahler manifolds, except $CP_{n-1,
1}$.

As pointed out in ref. [7] one should be able to derive systematically
the F function from the embedding of $\cal S$ into $Sp(2\, dim{\cal S} +2, R
)$. The action of the $\cal S$ isometry group $G$ of $\cal S$
on the section $\Omega$, under the
appropriate symplectic embedding, induces the right duality
transformations [9, 10] on the section, giving a recipe to calculate
$\partial_\Lambda F$ as a function of $X$, and reducing the problem to
the solution of an ordinary first--order differential equation.
In this letter we show that, following this procedure, we get, in a
rigorous way, a symplectic section $\Omega$ correspondig to the F
function (1.3).

Let $\Gamma$ denote the (target) space modular group of $\cal S$ [11,
12].
{}From the embedding of $\cal S$ into $Sp(2\, dim {\cal S} +2, R)$ we
retrieve the embedding of $\Gamma$ into  $Sp(2\, dim {\cal S} +2, Z)$.
Using the general formula proposed in ref. [8], we are able to construct
the $\Gamma$ automorphic function (the superpotential) for the case
in the title, writing it as a sum over integers describing a modular
lattice. This formula is explicitly recalled in the next section (see
eq. (2.11))
\vskip 0.5cm
\noindent
{\centerline{\bf 2. Construction of the section $\Omega$ for $\tisei$}}
\vskip 0.2cm
We start our programme by writing the coset representative
of the manifold $\tisei=
\o{G}{H}$ in projective coordinates [10]:
$$
M=\twomat {(1-ZZ^\dagger)^{- \o{1}{2}}} {(1-ZZ^\dagger)^{- \o{1}{2}}Z}
{Z^\dagger (1-ZZ^\dagger)^{- \o{1}{2}}} {1+ Z^\dagger (1-ZZ^\dagger)^{-
1}Z} \eqno(2.1)
$$
where $Z$ is a complex $3 \times 3$ matrix.
Let us denote by $A$ the $6 \times 3$ matrix given by
$$
A=((1-ZZ^\dagger)^{- \o{1}{2}}, (1-ZZ^\dagger)^{- \o{1}{2}} Z)\,,\eqno(2.2)
$$
where the indices of $A_i^I$ run as follows: $I=(i, i^*);\,\,
i, i^*=1,2,3$ ($i$ corresponds to the plus signs of the metric and $i^*$
to the minus signs).
Following the general procedure discussed in [7, 10] we
have to embed $G$ into the symplectic group of
dimension $(9+1) \times 2=20$. If we consider the isometry group
$G=SU(3,3)$, it is easily recognized that the three--index
antisymmetric representation of $G$ has the required
dimension. Hence let us define:
$$
t^{IJK}=\epsilon^{ijk} A_i^I A_j^J A_k^K \,. \eqno(2.3)
$$
The three index antisymmetric tensor
$t^{IJK}$ is acted on by the
matrix ${\cal B}= {\cal U}_I^{[I^\prime}
{\cal U}_J^{J^\prime}{\cal U}_K^{K^\prime] }$, where ${\cal U} \in SU(3,
3)$. One can verify that:

$${\cal U}^T C {\cal U}=C \,,\eqno (2.4)$$
where the matrix $C$ satisfies
$C^T=-C$, $C^2=-\un$, and can be viewed as acting on the triplet
$IJK$
as the Levi Civita symbol: $[Ct]^{IJL}=
\epsilon^{IJKLMN}t_{LMN}$. Moreover one has
$${\cal U}^\dagger E {\cal U}=E , \eqno(2.5)$$
where $E^2=\un$, $E^\dagger =E$ and where $E$ acts on the three--index
tensors
as $E=\eta_{I L^\prime}
\eta_{J M^\prime} \eta_{K N^\prime}$ ($\eta=(+,+,+,-,-,-)$), antisymmetrized
with respect to
$ L^\prime ,M^\prime, N^\prime$.

Equations (2.4) and (2.5)
show that $t^{IJK}$ realizes a symplectic embedding of $G$ into a
symplectic group of dimension $20$. Howewer, due to the signature of
the ``metric'' $E$ this group is $Usp(10,10)$ rather than $Sp(20,R)$.
The use of a
generalized Cayley matrix allows us to transform the $Usp(10,10)$
representation into the real symplectic one. Explicitly we set:
$$
T^{LMN}=
\epsilon^{ijk} A_i^I {\cal C}_I^L A_j^J {\cal C}_J^M A_k^K {\cal
C}_K^N=\hat{\cal C} t \,,\eqno(2.6)
$$
where
$$
{\cal C}={1 \over \sqrt{2}} \twomat {\un_3}{-i\un_3}{\un_3}{i\un_3}
$$
is the ``Cayley matrix'' in the $3 +3$ space (i.e. in the fundamental
representation of $SU(3,3)$) and $\hat{\cal C}$ as defined by equation
(2.6), is the generalized Cayley matrix in the 20 space (i.e. in the
three index antisymmetric representation of $SU(3,3)$). Note that both
$\cal C$ and $\hat{\cal C}$ are defined up to a phase.
The three--index representation is now written as (if we drop an
overall $1 \over \sqrt{2}$ factor)
$$
\eqalign{
T^{ijk} &=\hbox{det} \o{1+Z} {(1-ZZ^\dagger)^{1 \over 2}} \epsilon^{ijk}
\cr
T^{i^* ik} &=\epsilon^{ijk} \hbox{det} \o{1+Z} {(1-ZZ^\dagger)^{1 \over 2}}
\left (\o{i(Z-1)}{Z+1}\right )^{i^*}_i \cr
T^{i j^* k^*} &=\hbox{det} \o{1+Z} {(1-ZZ^\dagger)^{1 \over 2}}  \epsilon^{irs}
\left (\o{i(Z-1)}{Z+1}\right )^{j^*}_r \left (\o{i(Z-1)}{Z+1}\right )^{k^*}_s
\cr
T^{i^* j^* k^*} &=
\hbox{det} \o{1+Z} {(1-ZZ^\dagger)^{1 \over 2}} \hbox{det}\o{i(Z-1)}{Z+1}
\,,\cr}\eqno (2.7)
$$
where we adopt the convention: $\hbox{det} M_{3\times 3}=\o{1}{3!}
\epsilon^{ijk}
\epsilon_{i^*j^*k^*} M_i^{i^*}M_j^{j^*}M_k^{k^*}$.
Equation (2.7) gives the explicit form of the $Sp(20,R)$ section $\Omega$,
and each line of such an equation has to be interpreted as
$X^\Lambda$ or as $i\partial_\Lambda F(X)$. If we divide
by the overall factor $\hbox{det} (1+Z)$,
which gives, toghether with its antiholomorphic counterpart, a real
function contribution to the K\"ahler potential, we can read the first
$10$ components of $\Omega$ as the elements of the
matrix $X^{i^*}_i=[\o{i(Z-1)}{Z+1}]^{i^*}_i$ together with $X^0=1$.
By recalling the definition
$L^\Lambda=e^{\o {1}{2} G}
X^\Lambda$, and recalling
that for $\tisei$ the K\"ahler potential $G$ is expressed by [10]
$$
G(Z,Z^\dagger)=-log\, det(1-ZZ^\dagger)\, , \eqno (2.8)
$$
we obtain from (2.7)
the following identifications:
$$
\eqalign{
T^{ijk} &=\epsilon^{ijk} L^0 \cr
T^{i^*jk} &=\epsilon^{ijk}L^{i^*}_i \cr
T^{ij^*k^*} &=\epsilon^{i^* j^* k^*} \o {\partial}{\partial L_i^{i^*}}
\left(- \o{\hbox{det} L}{L^0}\right)=\epsilon^{i^* j^* k^*}
\o {\partial}{\partial L_i^{i^*}}
(i F(L)) \cr
T^{i^*j^*k^*} &=\epsilon^{i^* j^* k^*} \o {\partial}{\partial L^0}
\left(- \o{\hbox{det} L}{L^0}\right)=
\epsilon^{i^* j^* k^*} \o {\partial}{\partial L^0}
(i F(L))\, .\cr}\eqno(2.9)
$$
The last two eq.s (2.9) have to be interpreted as
differential equations satisfied by the F function. These are solved
by:
$$
F(L)=i \o{\hbox{det L}}{L^0}\eqno (2.10)
$$
(the same expression, of course, holds for $F(X)$).
If we introduce the ``special coordinates" $S^{i^*}_i=L^{i^*}_i/L^0
=X^{i^*}_i/X^0$,
we immediately recover the
standard expression of the K\"ahler
potential $G(S, \bar S)=-\hbox{log det} (S-\bar S)$, which is explicitly
obtained from (1.2) and which coincides with the
$-\hbox{log det} (1-
ZZ^\dagger)$, while posing $S=\o{i(Z-1)}{Z+1}$
(modulo the real part of a holomorphic function).

Our next step consists in searching an explicit formula for an
automorphic superpotential in the case of $\tis$. To this scope we
briefly recall
the general definition given in ref. [8]

$$ \hbox{log} \, || \Delta ||^2  \eq \hbox{log} \,
\left [ | \Delta  |^2  \, e^{G} \right ]=
\left [ \, - \, \sum_{(M_{\Sigma},N^{\Sigma})\in \Lambda_{\Gamma}}
\hbox{log} \o {|M_{\Sigma}X^{\Sigma} + i N^{\Sigma} \partial_{\Sigma} F |^2}
{X^{\Sigma}{\bar \partial}_{\Sigma} {\bar F} +
{\bar X}^{\Sigma}{\partial}_{\Sigma} F} \right ]_{reg}, \eqno(2.11)
$$
where the integers $(M,N)$ belong to a homogeneous lattice
$\Lambda_\Gamma$ associated with the target space modular group $\Gamma
\in Sp(2+2n, Z)$, where $\Gamma$ corresponds to a
suitable definition of $SU(3,3,Z
)$ (see next section).
The only difficult point, to make formula (2.11)
explicit, is to find the explicit
parametrization of the capital integers $(M,N)$ in terms of the
small integers $n$ spanning the Narain lattice for the $\tis$
orbifold.
In particular, for our case, we need the formula relating 20
``integers" $M^{IJK}$ in the three-times antisymmetric representation
of $Sp(20, R)$ (or equivalently $Usp(10,10)$) to the
``integers"  $l^I$ in the fundamental 6--dimensional
representation of $SU(3,3)$. We write ``integers" in quotes
because both $M^{IJK}$ and $l^I$ are not integers, rather they are
complex numbers parametrized by a double number of integers (real and
imaginary parts).
The solution of this problem is given by splitting the (complexified)
momentum lattice of $\tis$ into three conjugacy classes, and by
constructing the symplectic integers in terms of the integers belonging
to these classes. Let us give in some detail the analysis of the
momentum lattice and its behaviour under the modular group $SU(3,3,Z)$.
\vskip 0.5cm
\centerline{\bf 3. The momentum lattice of $\tis$ orbifold}
\centerline{\bf and the
modular group $SU(3,3,Z)$}
\vskip 0.2cm
Following a well--established literature we define a $T^{2n}/Z_N$
orbifold via a two--step process [13--15]. First we introduce a 2n
dimensional torus $T^{2n}$ by identifying points in $R^{2n}$ with
respect to the action of a lattice group $\Lambda_R$:
$$
X^\mu \sim X^\mu + v^\mu \quad ; \quad v^\mu \in \Lambda_R
\eqno (3.1)
$$
and we define $T^{2n}/Z_N$ by identifying points in $T^{2n}$ with
respect to the action of a point group $\cal P \sim Z_N$ that acts
cristallographically on the lattice $\Lambda_R$ and that is a subgroup
of $SO(2n)$:
$$
\eqalignno{
(\Theta X)^\mu &\sim X^\mu \quad ; \quad \Theta \in SO(2n)&(3.2a)\cr
(\Theta v)^\mu &\in \Lambda_R \,\,\, if \,\,\,v^\mu \in \Lambda_R\,. &(3.2b)
\cr}
$$
In the case of $\tis$, the standard choice of $\Lambda_R$
corresponds to $\Lambda_R=RA_2 \otimes RA_2 \otimes RA_2$, where
$RA_2$ is the root lattice of the simply laced Lie algebra $A_2$ [14].
In this way one easily obtains an $SO(6)$ rotation matrix $\Theta$,
which maps $\Lambda_R$ into itself and such that $\Theta^3=1$.

This construction has been discussed in the literature [13, 14] but we
need to recall it here. Indeed we have to illustrate the properties
of the momentum lattice we shall utilize in the derivation of the
$SU(3,3,Z)$ modular group and of the coefficients $M^{IJK}$. We begin
by introducing a complex structure in $R^6$. This is done by
substituting three complex coordinates $Z^i$ to the six real coordinates
$X^\mu$ via the relation:
$$
X^\mu=Z^i e_i^\mu +{\bar Z}^{i^*} e_{i^*}^\mu \eqno(3.3)
$$
($i,i^*=1,2,3$), where $\{e_i^{\mu}, e_{i^*}^\mu \}$ is a basis of six
complex, linear, independent vectors fulfilling the conditions [16]:
$$\eqalignno{
(e_i^\mu)^* &=e_{i^*}^\mu &(3.4a)\cr
(e_i,e_j)&=(e_{i^*},e_{j^*})=0 \quad (e_i,e_{j^*})=g_{ij^*}\,.&(3.4b)\cr}
$$
In (3.4) the scalar product $(\, ,\,)$ is defined with respect to some
constant metric $g_{\mu\nu}$ with (+,+,+,+,+,+) signature:
$$
(v,w)=v^\mu w^\nu g_{\mu\nu}\,. \eqno(3.5)
$$
The Hermitian form:
$$
g_{ij^*}=(e_i,e_{j^*})=g_{j^*i} \eqno(3.6)
$$
defines a Hermitian metric in $R^6$ equipped with the complex
structure (3.3):
$$
g_{\mu \nu}X^\mu X^\nu=2Z^i{\bar Z}^{j^*}g_{ij^*}\,. \eqno (3.7)
$$
The torus $T^6$ is obtained by setting the following identification of
points in $R^6$ [14]:
$$
Z^i=Z^i +(n^i +\Theta m^i)\sqrt{2} \eqno(3.8)
$$
where $n^i$, $m^i \in Z$
$$
\Theta=e^{2\pi i/3}\,. \eqno(3.9)
$$
Equation (3.8) corresponds to the modding by a lattice
$\Lambda_R=RA_2 \otimes RA_2 \otimes RA_2$, as claimed at the
beginning. Indeed for the algebra $A_2$ a system of simple roots is
given by the two--dimensional vectors:
$$
\alpha_1=(\sqrt{2},0)=\sqrt{2} \quad \alpha_2=\left(-\o{\sqd}{2},
\sqrt{\o{3}{2}}\right)=\sqd e^{2 \pi i /3}\,, \eqno(3.10)
$$
so that an element of the root lattice $RA_2$ can be represented by
the following complex number:
$$
n \alpha_1 +m\alpha_2=\sqd (n+m\Theta)\quad (n,m \in Z)\,.
\eqno(3.11)
$$
The dual--weight lattice $WA_2$ is spanned by the simple weights:
$$\eqalignno{
\lambda_1 &=\left(\o{1}{\sqd},\o{1}{\sqrt{6}}\right)
=\sqrt{\o{2}{3}}e^{\pi i/3}&(3.
12a)\cr
\lambda_2 &=\left(0,\sqrt{\o{2}{3}}\right)=\sqrt{\o{2}{3}}e^{\pi i/2}&(3.
12b)\cr}
$$
and a generic element of this lattice is represented by the following
complex number:
$$
p \lambda_1 + q \lambda_2 =\sqrt{\o{2}{3}}(p\omega_1 +q\omega_2)\,,
\eqno(3.13)
$$
where $\omega_1=e^{\pi i/3}$ and $\omega_2=e^{\pi i/2}$.

The metric $g_{ij^*}$ defined by eq. (3.6) enters, together with an
antisymmetric two--form $B_{ij^*}$, the two--dimensional $\sigma$ model
action on the $T^6$ torus [12,15]:
$$
S=\int d^2 \xi \partial_\alpha Z^i \partial^\alpha {\bar Z}^{j^*}
(g_{ij^*} +B_{ij^*})\,.\eqno(3.14)
$$
The nine complex parameters encoded in the complex 3 $\times$ 3 matrix
$$
M_{ij^*}=g_{ij^*} +B_{ij^*}\eqno(3.15)
$$
parametrize the orbifold $\tis$ moduli space whose special K\"ahler
geometry we have described in the previous section. For a generic
$T^6$ torus we would have 36 moduli corresponding to an arbitrary $
g_{\mu \nu}$ metric and an arbitrary $B_{\mu \nu}$ two--form. On the
contrary, for the orbifold, we just have the freedom of choosing
$g_{ij^*}$ and $B_{ij^*}$, since the complex structure (3.3) cannot be
deformed. Indeed in addition to the identification (3.8) under the
lattice group $\Lambda_R$ we also have the identification under the
point group $Z_3$. The generator $\Theta$  of $Z_3$ acts on the
complex coordinates $Z^i$ as a multiplication by $\Theta$:
$$
Z^i \sim \Theta Z^i=e^{2 \pi i /3} Z^i\, . \eqno (3.16)
$$
Equations (3.8) and (3.16) are compatible just because $\Theta$ acts
cristallographically on the lattice $\Lambda_R$. Indeed:
$$
\Theta[\sqd (n^i +m^i)]=\sqd(n^{i^\prime} + \Theta m^{i^\prime})\,,
\eqno(3.17)
$$
where $n^{i^\prime}=-m^i$, $m^{i^\prime}=n^i-m^i$, wich follows from
$$
\Theta^2=e^{4 \pi i/3}=-1-\Theta\,. \eqno(3.18)
$$
The momentum lattice is introduced in the usual way by considering the
plane waves $\hbox{exp}(iP_\mu X^\mu)$ and demanding that they are
single--valued on the torus $T^6$ with the complex structure (3.3).
This implies:
$$
\eqalignno{
P_\mu &=g_{\mu \nu}(P_i e^{i\nu}+{\bar P}_{i^*}e^{i^*\nu})&(3.19a)\cr
P_i&=\sqrt{\o {2}{3}}(p_i \omega_1 +q_i \omega_2)\quad ; \quad (p_i, q_i \in
Z)\,,&(3.19b)\cr}
$$
where $\{e^{i \nu}, e^{i^* \nu}\}$ $i, i^*=1,2,3$ form the dual basis to
the basis (3.4).
Following a standard procedure the winding modes can be included into
the momentum lattice, which becomes the Lorentzian 12 dimensional
Narain lattice [13] $\Lambda_W$ with signature $g_{\tilde\mu
\tilde \nu}=diag(+,+,+,+,+,+,-,-,-,-,-,-)$. In complete analogy to Eq.s
(3.19), one writes:
$$
\eqalignno{
P_{\tilde \mu}&= g_{\tilde \mu \tilde\nu}(P_I e^{I\tilde \nu}+
{\bar P}_{I^*} e^{I^* \tilde\nu}) \,\,\, (I=1,\cdots 6)&(3.20a)\cr
P_I &=\sqrt{\o{2}{3}}(p_I \omega_1 +q_I \omega_2)\, \, (p_I,q_I \in Z)
\,,&(3.20b)
\cr}
$$
where $e^{I \tilde \nu},e^{I^* \tilde \nu}$ are the basis vectors of the
Narain lattice $\Lambda_W$, and $P_{\tilde \mu}$ its elements. We have:
$$
\eqalignno{
(e^I,e^J)&=0 \, \, (e^{I^*},e^{J^*})=0&(3.21a)\cr
(e^{I},e^{J^*})&=g^{IJ^*}\,.&(3.21b)\cr}
$$
The metric $g^{I J^*}$ is a Hermitian metric with signature $diag({+,+,
+,-,-,-})$; hence the sesquilinear form $v^\dagger g w$ is invariant
against the transformations of a group isomorphic to $SU(3,3)$. This
is the origin of the $SU(3,3)$ symmetry discussed in the previous
section. Its role is clarified by considering the level-- matching
condition in the Narain lattice [13] (i.e. the equality of the left
and right masses):
$$
0=P^{\tilde\mu} P^{\tilde \nu}g_{\tilde \mu \tilde \nu}=
\o{2}{3} g^{I J^*}(p_Ip_J +q_Iq_J +p_Iq_J)\,.\eqno(3.22)
$$
Equation (3.22) follows upon straightforward substitution of eq. (3.19) and
(3.13) into $P^{\tilde \mu} P^{\tilde \nu}g_{\tilde \mu \tilde \nu}$.
By means of a similarity transformation, the metric $g^{I J^*}$ could
be reduced to the standard $SU(3,3)$ metric $\eta^{I J^*}=
diag(+,+,+,-,-,-)$. Indeed there exists a non--singular 6 $\times$ 6
matrix $\Omega$ such that:
$$
g^{I J^*}= (\Omega^\dagger \eta \Omega)^{I J^*}\,.
\eqno(3.23)
$$
Consider now the matrix $S$ given by:
$${ S}=
\twomat{\o{1}{\sqd}\un}{0}{\o{1}{\sqrt{6}}\un}{\o{2}{\sqrt{6}}\un}\,,
\eqno(3.24)
$$
where $\un$ is the unit matrix in the six dimensions. Equation (3.22) can be
rewritten as follows:
$$
0=u^T g u +v^T g v \eqno(3.25)
$$
where
$$
\twovec{u}{v}={S}\twovec{p}{q}\,.\eqno(3.26)
$$
The quadratic form (3.25) is the standard $SO(6,6)$ invariant form.
The elements of $SO(6,6)$ have the generic form:
$$
{\cal A}=\twomat{A}{B}{C}{D}\eqno(3.27)
$$
where the $6\times 6$ blocks fulfil the following conditions:
$$
\eqalign{
A^T g A+C^T g C &=g\cr
A^T g B +C^T g D&=0 \cr
B^TgB +D^T g D&=g\,.\cr}\eqno(3.28)
$$
In the $(u,v)$ basis the $Z_3$ generator $\Theta$ is given by the
matrix:
$$
\Theta_{(u,v)}=\twomat{\hbox{cos}(\o{2 \pi}{3}) \un}{-\hbox{sin}
(\o{2 \pi}{3}) \un}
{\hbox{sin}(\o{2 \pi}{3}) \un}{\hbox{cos}(\o{2 \pi}{3}) \un}\,.\eqno(3.29)
$$
The normalizer of $Z_3$ in $SO(6,6)$ is the group $SU(3,3)$. It is
composed of those matrices that have the special form:
$$
{\cal A}=\twomat{A}{-B}{B}{A}\,. \eqno(3.30a)
$$
with:
$$
A^T g A +B^T g B=0 \quad ; \quad A^TgB=B^TgA \,,\eqno(3.30b)
$$
In the $(p,q)$ basis the $Z_3$ generator is integer--valued:
$$
\Theta_{(p,q)}={S}^{-1} \Theta_{(u,v)}{S}=
\twomat{-\un}{-\un}{\un}{0}\,,\eqno(3.31)
$$
a proof that $Z_3$ acts cristallographically also on the Narain
lattice. In the torus compactification the modular group is
$\Gamma(T^6)=SO(6,6,Z)$, namely the subgroup of $SO(6,6)$ that maps the
Narain lattice into itself. In the orbifold case the modular group
$\Gamma(T^6/Z_3)$ is the subgroup of of $SU(3,3)$ that maps the Narain
lattice into itself. We name this group $SU(3,3,Z)$ and we easily
identify its elements. In the $(p,q)$ basis an $SU(3,3)$ element is
obtained from eq (3.30) via conjugation with the matrix $S$. We
get:
$$
{\cal A}_{(p,q)}={S}^{-1} {\cal A}{S}=
\twomat{A-\o{B}{\sqrt{3}}}{-2\o{B}{\sqrt{3}}}{2\o{B}{\sqrt{3}}}
{A+\o{B}{\sqrt{3}}}=\twomat{H}{H-K}{-H+K}{K}\,,\eqno(3.32)
$$
where we have set:
$$
H=A-\o{B}{\sqrt{3}} \quad ; \quad K=A+\o{B}{\sqrt{3}}\,.\eqno(3.33)
$$
In terms of the blocks $H,K$ the conditions (3.30b) become:
$$
\eqalignno{
K^Tg&K=H^TgH&(3.34)\cr
H^T g& H +K^T g K -\o{1}{2} K^Tg H -\o{1}{2} H^T gK=g\,.&(3.35)\cr}
$$
The group $SU(3,3,Z)$ is obtained by demanding that $H,K$ should be
integer--valued: $H_{IJ}, K_{IJ}\in Z$.
Since $\hbox{det} {\cal A}_{(p,q)}=\hbox{det} {\cal A}=1$, this condition is
compatible with the group structure and $SU(3,3,Z)$ is well defined.
Equivalently we can say that the group $SU(3,3,Z)$ is composed by all
the pseudo--unitary $6 \times 6$ matrices $\cal U$:
$$
{\cal U}^\dagger g {\cal U}=g \quad ; \quad \hbox{det}{\cal U}=1\eqno(3.36)
$$
that have the special form:
$$
{\cal U}=\um(K+H) +i \o{\sqrt{3}}{2}(K-H)\,,\eqno(3.37)
$$
$K$, $H$ being integer--valued matrices. In this case eq.s (3.34)
and (3.35) follow
from insertion of (3.37) into (3.36). The matrices $\cal U$ have
the property that acting on complex vectors of the form:
$$
l^I=\o{1}{\sqrt{2}}p^I +\o{i}{\sqrt{6}}(p^I+2q^I) \, \, p^I,q^I\in Z
\eqno(3.38)
$$
map them into complex vectors of the same form. Equations (3.37), (3.38)
are the final parametrization of the $SU(3,3,Z)$ modular group and of
the Narain lattice for the $\tis$ orbifold. They are the starting
point for the construction of the $M^{IJK}$ coefficients appearing in
the automorphic superpotential formula.
\vskip 0.5cm
\centerline{\bf 4. Construction of the $M^{IJK}$ coefficients}
\vskip 0.2cm
Na\"ively $l^I$ are in the six of $SU(3,3)$ while $M^{IJK}$ are in the
three--times antisymmetric representation, where we are considering the
$Usp(10,10)$ symplectic group. The only possibility of constructing
$M^{IJK}$ out of the $l^I$ momenta would be:
$$
M^{IJK}=l^{[I}l^Jl^{K]}
$$
which unfortunately is zero! The way out of this riddle results from
the properties of the Narain weight lattice $\Lambda_W$ which, while
modded with respect to its root sublattice  $\Lambda_R \subset
\Lambda_W$, splits into three conjugacy classes that are separately
invariant under the action of $SU(3,3,Z)$. Working in the $(p,q)$
basis (related to the actual momenta $l^I$ via eq. (3.38)) we define
$\Lambda_R$ as the sublattice, where $p,q$ have the form:
$$
\eqalign{
p^I&=2n^I-m^I\cr
q^I&=2m^I-n^I\,,\cr}\eqno(4.1)
$$
where $n^I, m^I \in Z$. This definition is inspired by the relation
between simple roots and simple weights in the $A_2$ case:
$$
\alpha_1=2 \lambda_1 +\lambda_2 \quad ; \quad \alpha_2=-\lambda_1
+2 \lambda_2\,,\eqno(4.2)
$$
so that
$$
n\alpha_1 +m \alpha_2=(2n-m)\lambda_1+(2m-n)\lambda_2 \,.\eqno(4.3)
$$
Equation (4.1) is equivalent to the condition:
$$
\o{1}{3} (p^I-q^I)\in Z \eqno(4.4)
$$
or
$$
\eqalignno{
n^I&=\o{1}{3}(2p^I+q^I)\in Z&(4.5a)\cr
m^I&=\o{1}{3}(p^I+2q^I)\in Z&(4.5b)\cr}
$$
An important result is the following:\par\noindent
{\bf Lemma}: {\underbar{The modular group $SU(3,3,Z)$ maps the root sublattice
$\Lambda_R$ into itself}}.\par\noindent
This follows straightforwardly from eq.(3.32)
$$
\o {1}{3}(p^\prime -q^\prime)=\o{1}{3}[Hp +(H-K)q +(K-H)p-Kq]=
H \o{1}{3}(2p+q)-K\o{1}{3}(2q+p)\eqno(4.6)
$$
If condition (4.4) (implying (4.5)) is fulfilled by $(p,q)$, the same
condition is fulfilled by the transformed $(p^\prime,q^\prime)$.
We can now write the complete Narain lattice $\Lambda_W$ as the sum of
three sublattices
$$
\Lambda_W=\Lambda_0 +\Lambda_1 +\Lambda_2 \,,\eqno(4.7)
$$
where $\Lambda_0=\Lambda_R$ is the already defined root lattice while
$\Lambda_1$ and $\Lambda_2$ are defined below:\par\noindent
{\bf{Definition}}: Let $\alpha=1,2$. A vector $(p,q)\in
\Lambda_W$ belongs to $\Lambda_\alpha \subset \Lambda_W$ if and only if
there exists an integer--valued non--zero six--vector $x^I\in Z$ such
that:
$$
\left(\o{1}{3}(p^I-q^I) +\o{\alpha}{3}x^I \right)\in Z\,.\eqno(4.8)
$$
The reason why the above is a good definition and why (4.7) is a good
decomposition is the following. For each value of the index $I$ the
difference $p^I-q^I$ can be $0,1,2$ mod $3$. The root lattice is that
sublattice such that $p^I-q^I=0$ mod $3$ for all the values of $I$.
$\Lambda_1$ is composed by those vectors such that $p^I-q^I=1$ mod $3$
for some values $I$ (at least one value) and $p^I-q^I=0$ mod $3$
in all the other cases. An analogous definition is given for
$\Lambda_2$. Since we have exhausted all the possibilities, any vector
$(p,q)\in \Lambda_W$ can be written as the sum of a vector in
$\Lambda_0$ plus a vector in $\Lambda_1$, plus a vector in $\Lambda_2$.
We have now the following:\par\noindent
\vfill\eject\noindent
{\bf{Theorem}}:
\par\noindent
\underbar{The lattices $\Lambda_\alpha$ are invariant
under the action of the modular group} \-$SU(3,3,Z)$.
\vskip 0.1cm \noindent
{\underbar{proof}}: Using the definitions (4.5) for each $(p,q)\in
\Lambda_\alpha$ we can write:
$$
\eqalignno{
n^I&=\o{1}{3}(2p^I+q^I)={\bar n}^I +\o{\alpha}{3}x^I&(4.9a)\cr
m^I&=\o{1}{3}(2q^I +p^I)={\bar m}^I- \o{\alpha}{3}x^I\,,&(4.9b)\cr}
$$
where ${\bar n}^I {\bar m}^I \in Z$. Under the action of $SU(3,3,Z)$
we get:
$$
\o{1}{3}(p^\prime -q^\prime)=Hn-Km=H{\bar n}-{\bar m}
+\o{\alpha}{3}(H+K)x \,;\eqno(4.10)
$$
since $H\bar n-K\bar m\in Z$ it follows that:
$$
\o{1}{3} (p^\prime-q^\prime) +\o{\alpha}{3}x^\prime \eqno(4.11)
$$
where
$$
x^\prime=-(H+K)x \,.\eqno(4.12)
$$
Therefore, provided $x^\prime\ne 0$, the image of a vector in
$\Lambda_\alpha$ is still in $\Lambda_\alpha$. On the other hand
$x^\prime$ cannot be zero. Indeed if $x^\prime$ were zero then the
image of $(p,q) \in \Lambda_\alpha$, under the $SU(3,3,Z)$ group
element $\gamma$ we consider, would be in $\Lambda_0$. Consider now the
inverse group element $\gamma^{-1}$: we obtain $\gamma^{-1} \gamma
(p,q)=(p,q) \in \Lambda_\alpha$, with $\alpha=1,2$. This would imply that
the image of the $\Lambda_0$ element $\gamma (p,q)$ under the $SU(3,3,
Z)$ transformation $\gamma^{-1}$ is not in $\Lambda_0$, contrary to the
lemma we have shown. Hence $x^\prime\ne 0$ and the theorem is proved.

Relying on this theorem we can now conclude the construction of the
$M^{IJK}$ coefficients. Extending the index $\alpha$ to the value $
\alpha=0$ corresponding to the root sublattice we can set:
$$
M^{IJK}=\epsilon^{\alpha \beta\gamma} l^I_\alpha
 l^J_\beta l^K_\gamma\,,\eqno(4.13)
$$
where the $l^I_\alpha \in \Lambda_\alpha$ is given in terms of $p^I,
q^I$ by eq. (3.38). The final formula for the automorphic
superpotential (where we are considering the $Usp(10,10)$
representation) is encoded in the following
$\zeta$--function regularization:
$$\eqalign{
\hbox{log} |\Delta|^2 e^G &= -\lim_{s \to 0}\, \o{d}{ds} \zeta (s) \cr
\zeta(s) &=\o{1}{\Gamma (s)} \int_0^\infty dt t^{s-1}
\sum_{l^I_\alpha\in\Lambda_\alpha}
e^{-it |M^{IJK} t_{IJK}|^2} \cr} \eqno(4.14)
$$
As can be seen, by summing independently on the three sublattices
$\Lambda_\alpha$ we are actually summing on $\Lambda_W$. The
coefficients (4.13) transform as a three--index antisymmetric
representation of $SU(3,3,Z)$ because of our theorem. Utilizing the
embedding of $SU(3,3)$ into $Usp(10,10)$ (or via Cayley in
$Sp(20,R)$) we discussed in section 2 we also see that $SU(3,3,Z)$
is a suitable discrete subgroup of $Usp(10,10)$ and that $M^{IJK}$
span the corresponding 20 dimensional symplectic representation of
this modular group.
\vskip 0.5cm
{\centerline {\bf 5. Conclusions}}
\vskip 0.2cm
In this paper we studied the special K\"ahler geometry of the
manifold $\tis=SU(3,3)/SU(3) \otimes SU(3) \otimes  U(1)$ utilizing
the symplectic embedding technique introduced in a previous
publication [7]. In this way we retrieved the cubic prepotential
$F(X)/(X^0)^2$ already discussed in the literature. Next we studied
the target space duality group of the
$\tis$ orbifold. This group acts as a discrete isometry on the
special K\"ahler moduli space and is isomorphic to a discrete
subgroup of $SU(3,3)$ which we called $SU(3,3,Z)$. Therefore the
moduli space of the $\tis$ orbifold is the special K\"ahler orbifold
\-${SU(3,3)/SU(3) \otimes SU(3) \otimes  U(1)}/SU(3,3,Z)$.\par\noindent
The group $SU(3,3,Z)$
acts in a $Usp(10,10)$ symplectic way on the orbifold lattice, whose
momenta and winding numbers can be classified in the threefold
antisymmetric representation of $SU(3,3,Z)$ where the three different
conjugacy classes of this discrete group are involved in the tensor
product. This symplectic action is crucial to define a duality--invariant
automorphic function via a $\zeta$ function regularization
of the determinant of a ``mass operator'' on the $\tis$ orbifold [7,8].
\vfill\eject
\centerline{\bf References}
\vskip 0.2cm

\def\title#1{~~{#1}~~}
\def\jour#1#2#3#4{ #1 $ {\underline {#2}}$ $(#3)~#4$}

\item{[1]}{E. Cremmer and
A. Van Proeyen, Class. and Quantum Grav. {$\underline{2}$} (1985) 445.}

\item{[2]}{E. Cremmer, C. Kounnas, A. Van Proeyen, J.P. Derendinger,
S. Ferrara, B. de Wit and L. Girardello, Nucl. Phys.
{$\underline{B250}$}  (1985)
385.}
\item{[3]}{B. de Wit and A. Van Proeyen, Nucl. Phys.
{$\underline{B245}$} (1984) 89.}
\item{[4]}{A. Strominger, Commun. Math. Phys. {$\underline{133}$} (1990) 163.}
\item{[5]}{L. Castellani, R. D'Auria and S. Ferrara, Phys. Lett.
{$\underline{B241}$} (1990) 57 and Class. and Quantum Grav.
{$\underline{1}$} (1990)
1767.}
\item{[6]}{R. D'Auria, S. Ferrara and P. Fr\`e, \jour{Nucl
Phys}{B359}{1991}{705}.}
\item{[7]}{P. Fr\`e and P. Soriani, Preprint SISSA 90/91/EP, to appear in
Nucl Phys B.}
\item{[8]}{S. Ferrara, C. Kounnas, D. L\"ust and F. Zwirner,
\jour{Nucl. Phys.}{B365}{1991}{431}.}
\item{[9]}{M.K, Gaillard and B. Zumino, \jour{Nucl. Phys.
}{B193}{1981}{221}.}
\item{[10]}{For a review see chapters IV.6 and I.6 in:
L. Castellani, R. D'Auria, P. Fr\'e \title{Supergravity and
superstrings: a geometric prospective}  World Scientific, Singapore
(1991).}
\item{[11]}{K.Kikkawa and M. Yamasaki, \jour{Phys. Lett.
}{149B}{1984}{357}.}
\item{}{N. Sakai and L. Senda, \jour{Progr. Theor. Phys.}{75}{1986}{692}.}
\item{}{V.P. Nair, A. Shapere, A. Strominger and F. Wilczek, Nucl.
Phys. {$\underline{B287}$} (1987) 402.}
\item{}{M. Dine, P. Huet and N. Seiberg, \jour{Nucl. Phys.}
{B322}{1989}{301}.}
\item{}{A. Shapere and F. Wilczek, \jour{Nucl. Phys.}{B320}{1989}{167}.}
\item{} {A. Giveon and M. Porrati, \jour{Phys. Lett.
}{246B}{1990}{54} and \jour{Nucl. Phys.}{355}{1991}{422}.}
\item{}{J. Lauer, J. Mas and H.P. Nilles, {\jour{Phys. Lett.
}{B226}{1989}{251} and \jour{Nucl Phys}{B351}{1991}{353}.}
\item{}{W. Lerche, D. L\"ust and N.P. Warner \jour{Phys.
Lett.}{B231}{1989}{417}.}
\item{}{M. Duff \jour{Nucl. Phys.}{B335}{1990}{610}.}
\item{}{A. Giveon, N. Malkin and E. Rabinovici, \jour{Phys.
Lett.}{B238}{1990}{57}.}
\item{}{J. Erler, D. Jungnickel and H.P. Nilles, Preprint MPI-Ph/91-90
 (1991).}
\item{}{S. Ferrara, D. L\"ust, A. Shapere and S. Theisen, \jour{
Phys. Lett.} {B233}{1989}{147}. }
\item{}{ S. Ferrara, D. L\"ust  and
S. Theisen,
\jour{Phys. Lett.}{B233}{1989}{147} and \jour{Phys. Lett.}{B242}{1990}{39}.}
\item{}{J. Schwartz, Calthech preprints CALT-6S-1581 (1990),
CALT-68-1728 (1991), CALT-68-1740 (1991)}
\item{}{J. Erler, D. Jungnickel and H.P. Nilles, Preprint MPI-Ph/91-81
(1991).}
\item{[12]}{A. Giveon, E. Rabinovici and G. Veneziano, \jour{Nucl.
Phys.}{B322}{1989}{167}.}
\item{[13]}{K.S. Narain, \jour{Phys. Lett.}{B169}{1986}{369}.}
\item{} {K.S. Narain, M.H. Sarmadi and E. Witten, \jour{Nucl.
Phys.}{B361}{1987}{414}.}
\item{} {L.J. Dixon, lectures given at 1987 ICTP Summer Workshop in
High Energy Physics and Cosmology}
\item{[14]}{D. Shevitz, \jour{Nucl. Phys.}{B338}{1990}{283}.}
\item{[15]}{E. Witten, \jour{Phys. Lett.} {B155}{1985}{151}.}
\item{}{S. Ferrara, C. Kounnas and M. Porrati, \jour{Phys. Lett.
}{B181}{1986}{263}.}
\item{} {M. Cvetic, J. Louis and B. Ovrut, \jour{Phys. Lett.}{B206}
{1988}{227}.}
\item{}{S. Ferrara and M. Porrati, \jour{Phys. Lett.}{B216}{1989}{289}.}
\item{}{M. Cvetic, J. Molera and B. Ovrut, \jour{Phys.
Rev.}{D40}{1989}{1140}.}
\item{[16]}{W. Lerche, A. Schellekens and N. Warner, \jour{Phys.
Rep.}{177}{1989}{1}.}
\vfill\eject\noindent
\nopagenumbers
\hskip 10cm \vbox{\hbox{CERN--TH 6364}
\hbox{SISSA 5/92/EP}}
\vskip 1.0cm
\centerline{\bf ON THE MODULI SPACE OF THE $T^6/Z_3$ ORBIFOLD}
\vskip 0.2cm
\centerline{\bf AND ITS MODULAR GROUP
\footnote*{\it Work
supported in part by Ministero dell'Universit\`a e
della Ricerca Scientifica e Tecnologica and by the Department of
Energy of the USA under contract DOE--AT03--88ER40 384, TASK E
}}

\vskip 1.5cm
\centerline{\bf Sergio Ferrara}
\vskip 0.2cm
\centerline{\sl CERN - Theory Division}
\centerline{\sl CH - 1211 Geneva 23, Switzerland}
\vskip 0.2cm
\centerline{\bf Pietro Fr\`e and Paolo Soriani}
\vskip 0.2cm
\centerline{\sl SISSA - International School for Advanced Studies}
\centerline{\sl Via Beirut 2, I-34100 Trieste, Italy}
\vskip 0.1cm
\centerline{ and}
\vskip 0.1cm
\centerline{\sl I.N.F.N. sezione di Trieste }
\centerline{\sl Area di Ricerca}
\centerline{\sl Padriciano 99, 34012, Trieste}
\vskip 1.0cm

\def\tisei{{\cal M}_{3.3}}
\def\tis{T^6/Z_3}
\centerline{\bf ABSTRACT}
\vskip 0.4truecm
We describe the duality group $\Gamma=SU(3,3,Z)$
for the Narain lattice of the $\tis$
orbifold and its action on the corresponding moduli space
${\tisei}/\Gamma$, where $\tisei=\o{SU(3,3)}{SU(3) \otimes SU(3)
\otimes  U(1)}$.
A symplectic embedding of the momenta and winding numbers allows
us to connect the orbifold lattice to the special geometry of
$\tisei$. As an application, a formal expression for an automorphic
function, which is a candidate for a non--perturbative superpotential,
is given.
\vskip 2.5truecm
\vbox{\hbox{CERN--TH 6364}
\hbox{SISSA 5/92/EP}
\hbox{January 1992}}

\bye